\documentclass[trackchanges,twocolumn]{aastex701}
\usepackage{lipsum}
\usepackage{amsmath}
\usepackage{enumitem}
\usepackage{booktabs}

\begin{document}

\title{Bar Formation During a Gaia-Sausage-Enceladus-like Merger Event}

\author[0000-0001-8962-663X]{Bin-Hui Chen}
\affiliation{Tsung-Dao Lee Institute, Shanghai Jiao Tong University,
Shanghai 200240, People's Republic of China}
\affiliation{Department of Astronomy, School of Physics and
  Astronomy, Shanghai Jiao Tong University, 800 Dongchuan Road,
Shanghai 200240, People's Republic of China}
\affiliation{State Key Laboratory of Dark Matter Physics, School of
  Physics and Astronomy, Shanghai Jiao Tong University, Shanghai
200240, People's Republic of China}
\affiliation{Key Laboratory for Particle Astrophysics and Cosmology
  (MOE) / Shanghai Key Laboratory for Particle Physics and Cosmology,
Shanghai 200240, People's Republic of China}
\affiliation{LIRA, Observatoire de Paris, Université PSL, CNRS, 92195
Meudon, France}
\email{2000cbh@sjtu.edu.cn}

\author[0000-0001-5604-1643]{Juntai Shen}
\affiliation{Department of Astronomy, School of Physics and
  Astronomy, Shanghai Jiao Tong University, 800 Dongchuan Road,
Shanghai 200240, People's Republic of China}
\affiliation{State Key Laboratory of Dark Matter Physics, School of
  Physics and Astronomy, Shanghai Jiao Tong University, Shanghai
200240, People's Republic of China}
\affiliation{Key Laboratory for Particle Astrophysics and Cosmology
  (MOE) / Shanghai Key Laboratory for Particle Physics and Cosmology,
Shanghai 200240, People's Republic of China}
\email[show]{jtshen@sjtu.edu.cn}
\email{jtshen@sjtu.edu.cn}
\correspondingauthor{Juntai Shen}

\author[0000-0002-5213-4807]{Paola Di Matteo}
\affiliation{LIRA, Observatoire de Paris, Université PSL, CNRS, 92195
Meudon, France}
\email{paola.dimatteo@obspm.fr}

\begin{abstract}

  Bars are among the most prominent galactic structures, yet their
  formation mechanisms remain incompletely understood. They can form
  either internally, via dynamical instabilities, or externally,
  triggered by interactions with other galaxies. The impact of
  mergers on bar formation and survival, however, has not been
  thoroughly investigated. To explore the influence of mergers on
  bars, we construct a suite of \textit{N}-body merger pairs where a
  Gaia-Sausage-Enceladus-like radially biased satellite disk galaxy
  merges with a central disk galaxy during its bar formation. With
  the central galaxy fixed, the satellite varies in merger
  parameters: the mass ratio $m/M$ relative to the central galaxy,
  the impact parameter $b$, and the orbital inclination angle
  $\theta_i$ relative to the central disk. We find that the bar
  survival probability decreases with increasing $m/M$. Mergers with
  $m/M\lesssim1/10$ generally preserve the forming bar, whereas those
  with ${m/M}\geq1/2$ tend to destroy it, producing more
  early-type-like remnants. For intermediate mass ratios ($1/5 \leq
  m/M \leq 1/3$), several models yield ``weakening bars'', in which
  the bar survives the merger but gradually decays during subsequent
  secular evolution, possibly due to interactions between nested
  double bars formed from merger debris. In contrast to $m/M$, $b$
  and $\theta_i$ have only secondary and stochastic effects on bar
  survival. The different influences of these three merger parameters
  can be naturally explained by the tidal force exerted by the
  satellite on the forming bar, which tends to weaken the bar when
  the satellite crosses it nearly perpendicular to its major axis.

\end{abstract}


\section{Introduction}

Bars are known to be prevalent in disk galaxies in the local universe
\citep[e.g.,][]{eskrid_etal_2000, menend_etal_2007, barazz_etal_2008,
sheth_etal_2012, simmon_etal_2014, erwin_2018, lee_etal_2019}, and
the unprecedented capabilities of JWST have recently revealed many
barred galaxies at high redshift as well
\citep[e.g.,][]{costan_etal_2023, guo_etal_2023, lecont_etal_2024}.
Bars play a critical role in shaping their host galaxies, including
driving corotation and Lindblad resonances
\citep[e.g.,][]{dehnen_2000, haywoo_etal_2024}, redistributing mass
distribution \citep[e.g.,][]{sellwo_2014, debatt_etal_2017,
berald_etal_2023, chen_etal_2024}, funneling gas toward galactic
centers to feed central black holes \citep[e.g.,][]{athana_1992,
kor_ken_2004, li_etal_2016, li_etal_2017, kat_viv_2024}, exchanging
angular momentum with the dark matter halo and heating stellar disks
\citep[e.g.,][]{athana_2002, sel_bin_2002, athana_2003_a,
athana_2003_b, min_fam_2010, long_etal_2014}, contributing to the
development of complex multi-populations of galaxies
\citep[e.g.,][]{debatt_etal_2020, che_li_2022}, and so on. A
comprehensive understanding of bar formation and evolution is
therefore essential for advancing our knowledge of galactic physics.

The mechanism of bar structure formation remains an open question.
Broadly, there are two formation pathways: internal formation,
arising from dynamical instabilities of the galactic disk
\citep[e.g.,][]{toomre_1981, sellwo_1981, sellwo_2016,
  fujii_etal_2018, binney_2020, bland_etal_2023, che_she_2025,
chen_etal_2025}, and externally driven or promoted formation,
triggered by interactions with neighboring galaxies
\citep[e.g.,][]{noguch_1981, gerin_etal_1990, miw_nog_1998,
lokas_etal_2014, zhe_she_2025, zheng_etal_2025}. For the latter, the
physical effects of a fly-by or merging satellite remain incompletely
understood. Some studies suggest that tidal interactions may either
facilitate or delay bar formation \citep{lokas_etal_2016,
moetaz_etal_2017, zana_etal_2018, pet_wad_2018, lokas_etal_2018};
\citet{ghosh_etal_2021} found that a minor merger can weaken an
existing bar in some idealized \textit{N}-body smooth particle
hydrodynamics simulations, whereas \citet{merrow_etal_2024} reported
that a similar merger in one of the Auriga cosmological zoom-in
simulations \citep{grand_etal_2017, grand_etal_2024} can promote bar
formation in a Milky Way–like galaxy.

As a typical disk galaxy, the Milky Way (MW) is known to host a bar
structure \citep{ger_vie_1986, binney_etal_1991, shen_etal_2010,
nes_lan_2016, portai_etal_2017}. With the advent of high-precision
astrometric data from the \textit{Gaia} mission \citep{gaia_dr1,
gaia_dr2, gaia_edr3, gaia_dr3}, \citet{beloku_etal_2018} and
\citet{helmi_etal_2018} identified a population of inner halo stars
with orbital and chemical properties distinct from the bulk of the
Galactic disk (in hindsight, see also \citealt{nis_sch_2010} for the
identification of such stars before the Gaia era). The similarity in
age and chemical abundance between such halo stars and those in
external dwarf galaxies suggests an \textit{ex situ} origin from a
single ancient accreted satellite of these halo stars (e.g.,
  \citealt{haywoo_etal_2018, gallar_etal_2019, conroy_etal_2019_b,
das_etal_2020, feuill_etal_2020, bonaca_etal_2020}). This accretion
event, now widely recognized as the last major merger in the Galactic
assembly history, is commonly referred to as the
Gaia-Sausage-Enceladus (GSE) event. Since the seminal studies by
\citet{beloku_etal_2018} and \citet{helmi_etal_2018}, the GSE has
become a central topic in Galactic archaeology
\citep[e.g.,][]{myeong_etal_2019, koppel_etal_2019, beloku_etal_2020,
  naidu_etal_2020,xiang_etal_2022, khoper_etal_2023, deng_etal_2024,
liu_etal_2025}.

Therefore, as a barred galaxy that has experienced a significant
merger, the Milky Way serves as a valuable test ground for
investigating the relationship between galactic mergers and bar
formation. Depending on the timing relative to bar formation, merger
events can be categorized as pre-bar-formation, during-bar-formation,
or post-bar-formation. Despite debates
\citep[e.g.,][]{nepal_etal_2024}, many studies suggest that both the
formation of the Galactic bar \citep{bovy_etal_2019, wylie_etal_2022,
sander_etal_2024, haywoo_etal_2024} and the GSE event
\citep{beloku_etal_2018, helmi_etal_2018, haywoo_etal_2018,
dimatt_etal_2019} occurred approximately $8\sim10\ \mathrm{Gyr}$ ago,
implying that the GSE might be a during-bar-formation merger event.
However, whether the GSE event is responsible for the formation of
the Galactic bar, or for promoting or suppressing it, remains poorly
understood. As such, understanding galactic mergers, particularly
those coinciding with the bar formation epoch, is critical for both
constraining externally driven bar formation mechanisms and advancing
Galactic archaeology. For instance, \citet{naidu_etal_2021}
constrained the properties of the GSE event by comparing simulations
with observed kinematics and chemistry of GSE stars in the H3
Spectroscopic Survey \citep{conroy_etal_2019}.

In this study, we study the influence of during-bar-formation mergers
on bar formation. We construct a suite of \textit{N}-body merging
pairs consisting of two similarly configured disk galaxies, possibly
with different sizes, in which a satellite merges with a central
galaxy while the latter is undergoing internal bar formation. With
the same central galaxy, we vary the satellite with three merger
parameters: (1) the satellite-to-central galaxy mass ratio $m/M$; (2)
the impact parameter $b$ of the satellite's incident velocity; and
(3) the inclination angle $\theta_i$ between the merging orbital
plane and the central galaxy's disk plane. With these models, we
systematically investigate how a GSE-like merger
event---characterized by a radially biased orbit and spanning the
plausible range of mass ratios inferred for the GSE event---affects
bar formation in the central galaxy.

The structure of this paper is as follows.
In Section~\ref{sec: models},
we describe the setup of the \textit{N}-body merging pairs.
Section~\ref{sec: results} presents the main results regarding bar
survival probabilities. In Section~\ref{sec: discussion}, we examine
the dynamical mechanisms shaping the bar survival probabilities and
discuss the bar weakening phenomenon. Finally, we summarize this
paper in Section~\ref{sec: summary}.

\begin{figure}
  \centering
  \includegraphics[width=.48\textwidth]{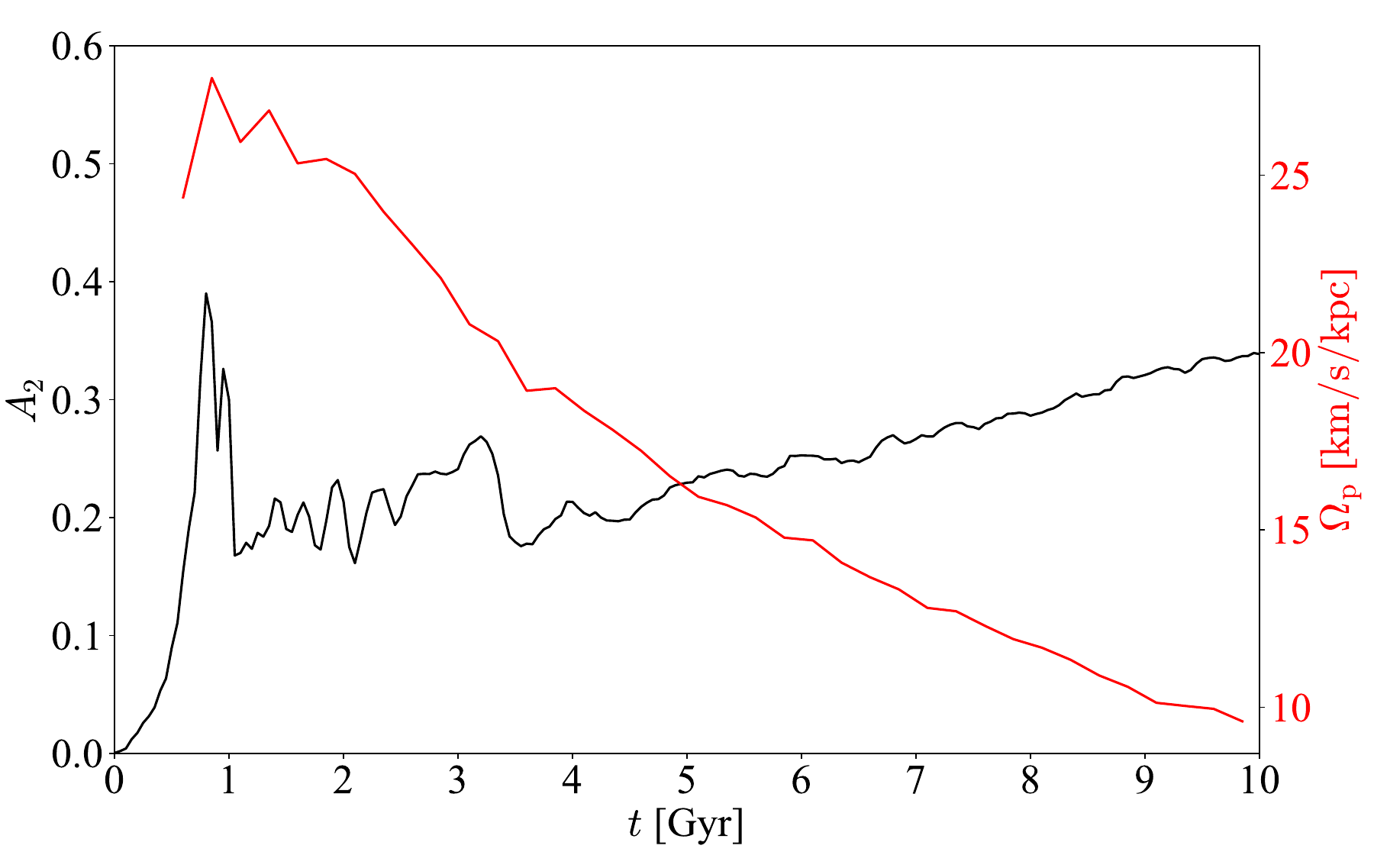}
  \caption{Temporal evolution of the bar strength $A_2$ (black) and
    pattern speed $\Omega_\mathrm{p}$ (red) for the central galaxy
    evolved in isolation. A prominent bar forms within
    $1\ \mathrm{Gyr}$ and subsequently undergoes secular growth
  accompanied by a corresponding decline in pattern speed.}
  \label{fig: isolated model}
\end{figure}

\begin{figure*}
  \centering
  \includegraphics[width=.81\textwidth]{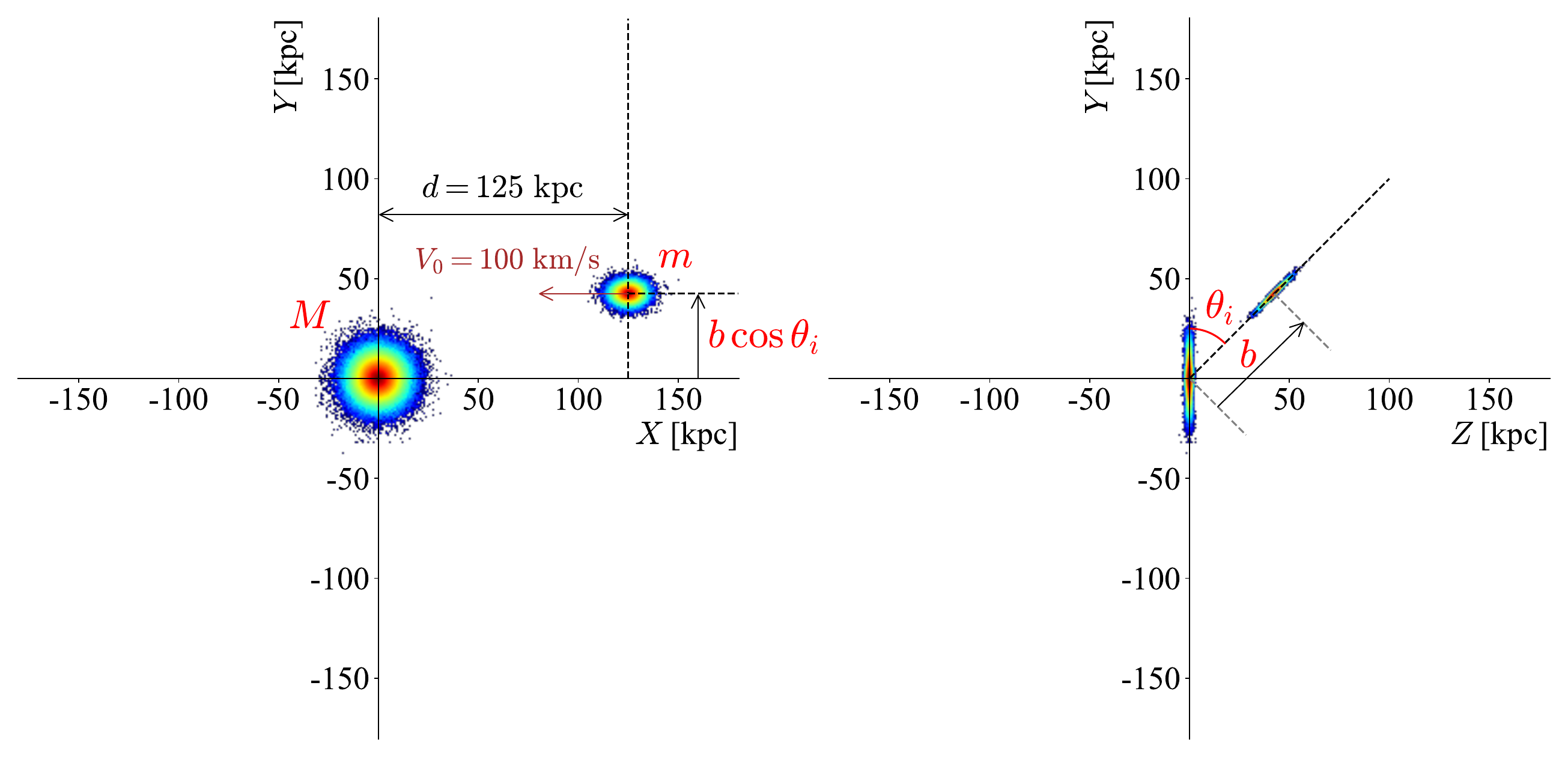}
  \caption{Illustration of the merger parameters with stereoscopic
    projections: the mass ratio $m/M$, impact parameter $b$, and
    orbital inclination angle $\theta_i$. The initial separation
    between the two galaxies is fixed as $d=125\ \mathrm{kpc}$ along
    the $X$-axis, while the positions in other directions vary
    according to $b$ and $\theta_i$. The satellite's incident velocity
    is fixed as $-100\ \mathrm{km/s}$ along the $X$-axis.
  }\label{fig: IC parameters}
\end{figure*}

\section{Models}\label{sec: models}

We construct our models using the \texttt{AGAMA} package
\citep{vasili_2019}. Each system consists of a central and a
satellite galaxy, both comprising a dark matter (DM) halo and a
stellar disk. The DM halo follows a Navarro–Frenk–White (NFW) density
profile \citep{navarr_etal_1995, navarr_etal_1996} with an outer cutoff:
\begin{equation}\label{eq: halo}
  \rho_\mathrm{NFW}(r)=\rho_0\left(\frac{r}{a}\right)^{-1}\left(1+\frac{r}{a}\right)^{-2}\times\exp{\left[-\left(\frac{r}{r_\mathrm{c}}\right)^2\right]},
\end{equation}
where $r$ is the spherical radius, $\rho_0$ is the normalization
constant set by the total halo mass, $a$ is the halo scale length,
and $r_\mathrm{c}$ is the outer truncation radius of the density profile.

The stellar disk is described by a quasi-isothermal distribution
function \citep{vasili_2019}:
\begin{equation}\label{eq: disk DF}
  \begin{aligned}
    f(\boldsymbol{J})&=\dfrac{\tilde{\Sigma}\Omega}{2\pi^2
    \kappa^2}\times\dfrac{\kappa}{\tilde{\sigma}_R^2}\exp\left(-\dfrac{\kappa
    J_R}{\tilde{\sigma}_R^2}\right) \\
    &\times \dfrac{\nu}{\tilde{\sigma}_z^2}\exp\left(-\dfrac{\nu
    J_z}{\tilde{\sigma}_z^2}\right)\times\left.\left\{
      \begin{array}{ll}1&\text{if
        }J_\phi\geq0,\\\exp\left(\dfrac{2\Omega
        J_\phi}{\tilde{\sigma}_R^2}\right)&\text{if }J_\phi<0,
      \end{array}\right.\right.
    \end{aligned}
  \end{equation}
  where $\kappa$/$\nu/\Omega$ is the epicycle/vertical/circular
  frequency, $J_R,\ J_\phi,\ J_z$ are the actions, and
  \begin{equation}
    \begin{split}
      \tilde{\Sigma}(R) &= \Sigma_0 \exp{(-R/R_\mathrm{d})}, \\
      \tilde{\sigma}_R^2(R) &= \sigma_{R,0}^2\exp{(-2R/R_{\sigma_R})}, \\
      \tilde{\sigma}_z^2 (R) &= 2 h_z^2 \nu^2(R),
    \end{split}
  \end{equation}
  where $R$ is the cylindrical radius, $R_\mathrm{d}$ is the disk
  scale length, $h_z$ is the vertical scale height, $\Sigma_0$ is the
  central surface density, $\sigma_{R,0}$ is the central radial
  velocity dispersion, and $R_{\sigma_R}$ is the scale length of the
  dispersion profile.

  \begin{table*}[htbp!]
    \centering
    \caption{Simulation Configuration: Galaxy Components and Merger
      Parameters\footnote{Part I lists the structural properties of the
    central galaxy, while Part II gives the merger parameters.}}
    \label{table: simulation parameters}
    \begin{tabular*}{\textwidth}{@{\extracolsep{\fill}} lcccccccccccc}
      \toprule
      \multicolumn{13}{l}{\textbf{Part I: Central Galaxy Parameters}} \\
      \midrule
      \textbf{Component} & $M$ [$10^{10} M_\odot$] & $N_\mathrm{p}$
      \footnote{Particle number.} &
      \multicolumn{2}{c}{$a$ or $R_\mathrm{d}$ [kpc]} &
      \multicolumn{2}{c}{$r_\mathrm{c}$ or $h_z$ [kpc]} &
      \multicolumn{2}{c}{$R_{\sigma_R}$ [kpc]} &
      \multicolumn{2}{c}{$\sigma_{R,0}$ [km/s]} &  \\
      \cmidrule(lr){1-1} \cmidrule(lr){2-2} \cmidrule(lr){3-3}
      \cmidrule(lr){4-5} \cmidrule(lr){6-7} \cmidrule(lr){8-9}
      \cmidrule(lr){10-11}
      DM halo      & 100.0 & 1,000,000 & \multicolumn{2}{c}{20.0} &
      \multicolumn{2}{c}{200.0} & \multicolumn{2}{c}{--} &
      \multicolumn{2}{c}{--} & \\
      Stellar disk & 5.0   & 1,000,000 & \multicolumn{2}{c}{2.5}  &
      \multicolumn{2}{c}{0.25}  & \multicolumn{2}{c}{5.0} &
      \multicolumn{2}{c}{100} & \\

      \midrule[1pt]
      \multicolumn{13}{l}{\textbf{Part II: Merger Parameters}} \\
      \midrule
      \textbf{Parameter} & \multicolumn{12}{c}{\textbf{Values}} \\
      \midrule
      $m/M$      & 1/1 & 1/2 & 1/3 & 1/4 & 1/5 & 1/6 & 1/8 & 1/10 &
      1/20 & & & \\
      $b/\mathrm{kpc}$ & -100 & -80 & -60 & -40 & -20 & 0 & 20 & 40 &
      60 & 80 & 100 & \\
      $\theta_i/\mathrm{deg}$ & 0 & 15 & 30 & 45 & 60 & 75 & 90 & & & & & \\
      \bottomrule
    \end{tabular*}
  \end{table*}

  Part I of Table~\ref{table: simulation parameters} lists the
  parameters of the central galaxy. This setup is typical for a cold
  disk model, which can form a bar spontaneously through bar
  instability \citep[e.g.,][]{shen_etal_2010, sel_ger_2020,
  tepper_etal_2021, bland_etal_2023, che_she_2025, chen_etal_2025}.

  We employ the following parameters to quantify the bar structure:
  \begin{equation*}
    \begin{aligned}
      &\text{Bar strength } A_2 = \left|\dfrac{\sum_{j} m_j
      \exp{(2i\phi_j)}}{\sum_j m_j}\right|, \\
      &\text{Bar pattern speed } \Omega_\mathrm{p} = \frac{\Delta
      \phi_2}{\Delta t},
    \end{aligned}
  \end{equation*}
  where the bar phase angle is defined as
  \[
    \phi_2 = \frac{1}{2} \arg\left(\dfrac{\sum_{j} m_j
    \exp{(2i\phi_j)}}{\sum_j m_j}\right).
  \]
  Here, $m_j$ and $\phi_j$ are the mass and cylindrical azimuthal
  angle of the $j$-th particle, respectively. Unless otherwise
  stated, the summations are taken over stellar particles with
  cylindrical radius $R < 15\ \mathrm{kpc}$.

  Figure~\ref{fig: isolated model} shows the temporal evolution of
  the bar strength and pattern speed for the central galaxy evolved
  in isolation (hereafter, the ``isolated model''). A prominent bar
  forms within $1\ \mathrm{Gyr}$. With the secular growth of bar
  strength, the pattern speed declines at a rate approximately
  $2\ \mathrm{km}\ \mathrm{s}^{-1}\ \mathrm{kpc}^{-1}\ \mathrm{Gyr}^{-1}$,
  which is about half the estimated value of the Milky Way
  \citep{chiba_etal_2021}.

  For the merger models, the satellite's parameters are scaled
  according to its mass ratio relative to the central galaxy:
  \begin{equation*}
    \begin{aligned}
      \text{Given } \gamma \equiv m/M,& \text{ then} \\
      m &= \gamma M , \\
      N_\mathrm{p}(\text{satellite}) &= \gamma N_\mathrm{p}(\text{central}), \\
      R_\mathrm{d}(\text{satellite}) &= \gamma^{1/3}
      R_\mathrm{d}(\text{central}), \\
      h_z(\text{satellite}) &= \gamma^{1/3} h_z(\text{central}), \\
      a(\text{satellite}) &= \gamma^{1/3} a(\text{central}), \\
      R_{\sigma_R}(\text{satellite}) &= \gamma^{1/3}
      R_{\sigma_R}(\text{central}), \\
      \sigma_{R,0}(\text{satellite}) &= \gamma^{1/3}
      \sigma_{R,0}(\text{central}).
    \end{aligned}
  \end{equation*}
  Here, the satellite's velocity dispersion is inferred from the
  Tully–Fisher relation \citep{tul_fis_1977}, which implies an
  approximate $\sigma \propto V_\mathrm{c} \propto M^{1/3}$ scaling
  in the $B$ band centered at $0.45\ \mu\mathrm{m}$
  \citep{bin_tre_2008}. Although the Tully–Fisher relation exhibits
  intrinsic scatter in its power-law index
  \citep[e.g.,][]{courte_1997, mcgaug_etal_2000}, this uncertainty
  has a negligible effect on the subject under study (see
  Section~\ref{sec: discussion: b and theta}).

  The initial configurations of the merging pairs are illustrated in
  Figure~\ref{fig: IC parameters}. We vary only the key merger
  parameters, as listed in Part II of
  Table~\ref{table: simulation parameters}:
  9 values of the merger mass ratio $m/M$; 11 values of
  the impact parameter $b$ (negative for retrograde orbits); and 7
  values of the inclination angle $\theta_i$. This results in a total
  of 693 models. For each model, the satellite is initially placed at
  $\mathbf{P}_0=(125\ \mathrm{kpc},\ b\cos\theta_i,\ b\sin\theta_i)$
  with an incident velocity
  $\mathbf{V}_0=(-100,\ 0,\ 0)\ \mathrm{km\,s^{-1}}$, which are
  chosen to ensure that the merger occurs on a radially biased orbit
  while the central galaxy is undergoing spontaneous bar formation.
  Note that, to reduce the dimensionality of the parameter space, we
  align the satellite's disk plane with its orbital plane, which has
  no impact on our main conclusions (see Section~\ref{sec: discussion}).

  We name these models as
  \texttt{mr$\{M/m\}$ip$\{b\}$ia$\{\theta_i\}$} based on their merger
  parameters; for example, model \texttt{mr20ip-60ia45} is the case
  with $m/M = 1/20$, $b = -60~\mathrm{kpc}$, and $\theta_i = 45^\circ$.

  Each model is evolved using the \texttt{GADGET4} code
  \citep{spring_etal_2021} for $10\ \mathrm{Gyr}$ to ensure that the
  merger process and subsequent dynamical evolution thoroughly
  saturate. In most cases, the satellite merges with the central
  galaxy within about $1.5\ \mathrm{Gyr}$ after 2–3 pericentric
  passages. After the merger, the system evolves in a secular manner.

  \begin{figure*}
    \centering
    \includegraphics[width=.98\textwidth]{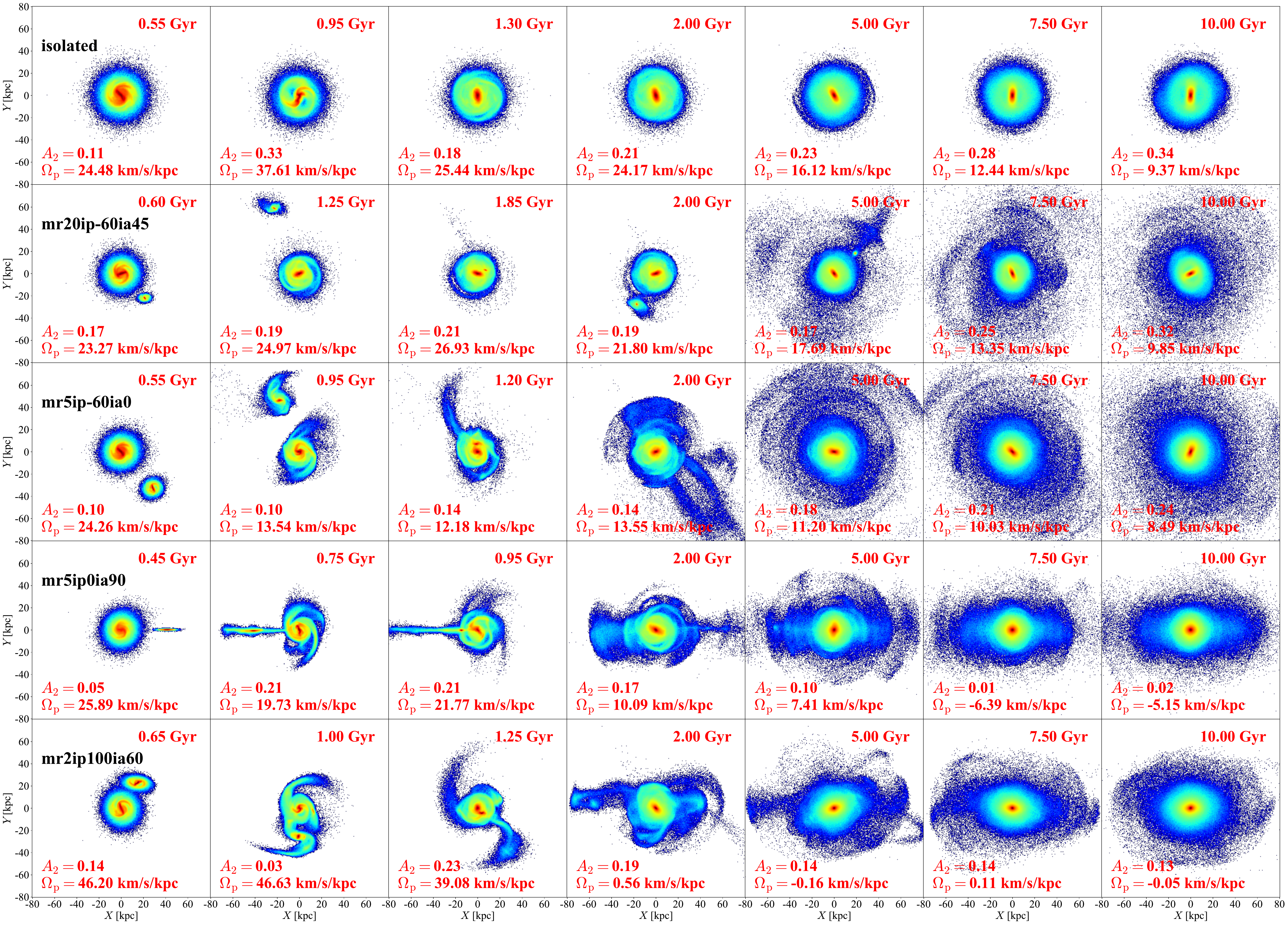}
    \caption{Face-on views of several representative models. The top
      row shows the isolated model. The second row depicts a minor
      merger, where the satellite's effect on the bar is negligible.
      The third row presents a model in which the bar survives in a
      non-trivial merger. The fourth row illustrates a weakening bar
      that survives the merger but gradually decays subsequently in
      secular evolution. The fifth row shows a merger-produced
      elliptical-like system with negligible rotation, resulting from a
      violent merger. The bar strengths and pattern speeds of the
      central galaxy are annotated in the lower-left corner of each
    panel.}\label{fig: snapshots}
  \end{figure*}

  \begin{figure}
    \centering
    \includegraphics[width=.47\textwidth]{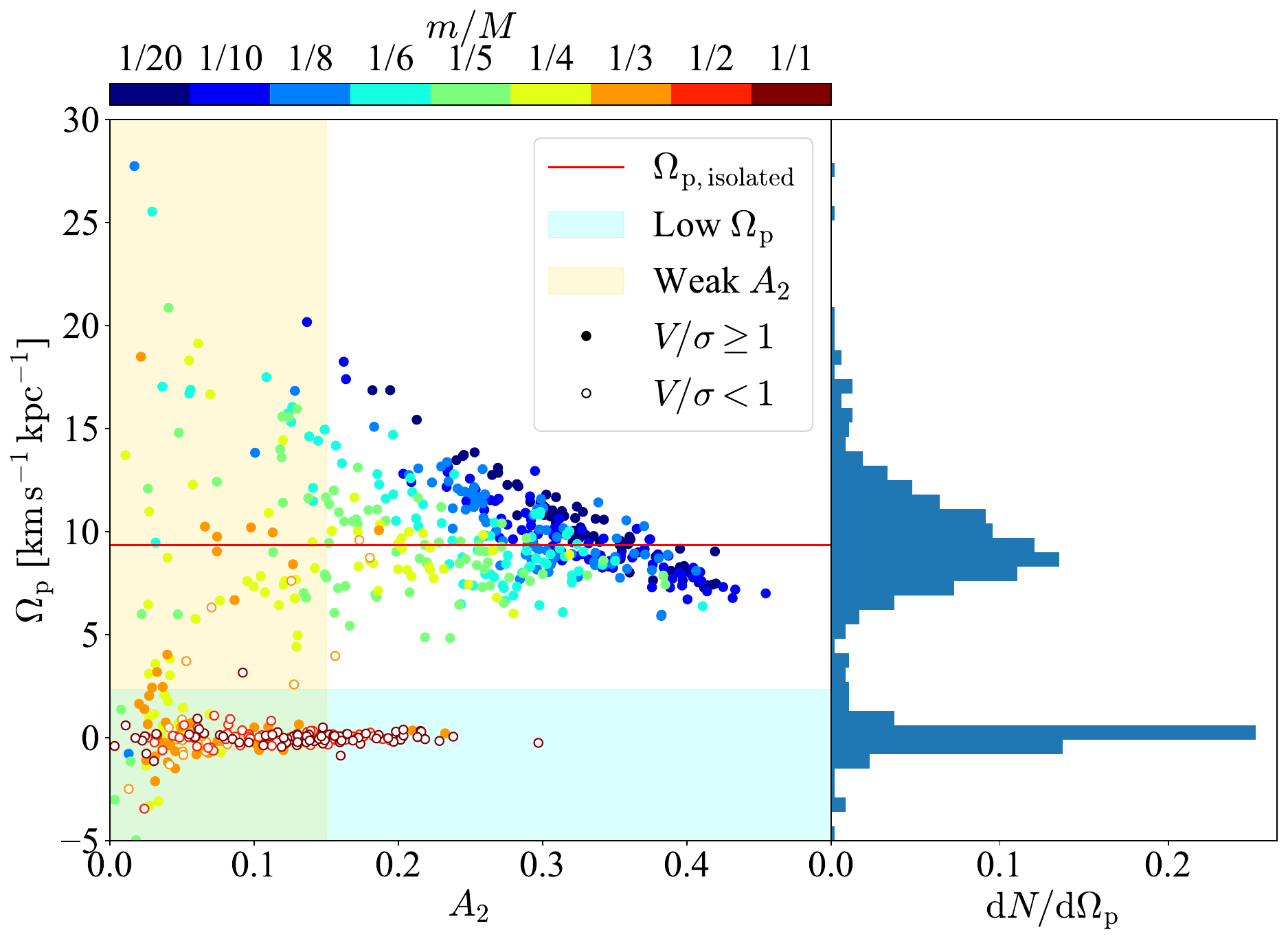}
    \caption{\textbf{Left}: Pattern speed $\Omega_\mathrm{p}$ as a
      function of bar strength $A_2$, with three models of very high
      pattern speed omitted for clarity (one with
        $\Omega_\mathrm{p}\sim40\ \mathrm{km\,s^{-1}\,kpc^{-1}}$ and
        two with
      $\Omega_\mathrm{p}\sim60\ \mathrm{km\,s^{-1}\,kpc^{-1}}$). The
      color of each point indicates the merger mass ratio. The red
      line marks the pattern speed of the isolated model,
      $\Omega_\mathrm{p, isolated}$. The cyan and yellow shaded
      regions denote low pattern speeds
      ($\Omega_\mathrm{p}<0.25\,\Omega_\mathrm{p, isolated}$) and
      weak bars ($A_2<0.15$), respectively. Models with an average
      azimuthal streaming velocity–to–dispersion ratio $V/\sigma<1$
      within $10<R/\mathrm{kpc}<20$ are shown as unfilled circles.
      \textbf{Right}: Histogram of $\Omega_\mathrm{p}$, showing one
      peak near $\Omega_\mathrm{p, isolated}$ and another near
      $\Omega_\mathrm{p}=0~\mathrm{km\,s^{-1}\,kpc^{-1}}$.
    }\label{fig: pattern speeds}
  \end{figure}

  \begin{figure*}
    \centering
    \includegraphics[width=.95\textwidth]{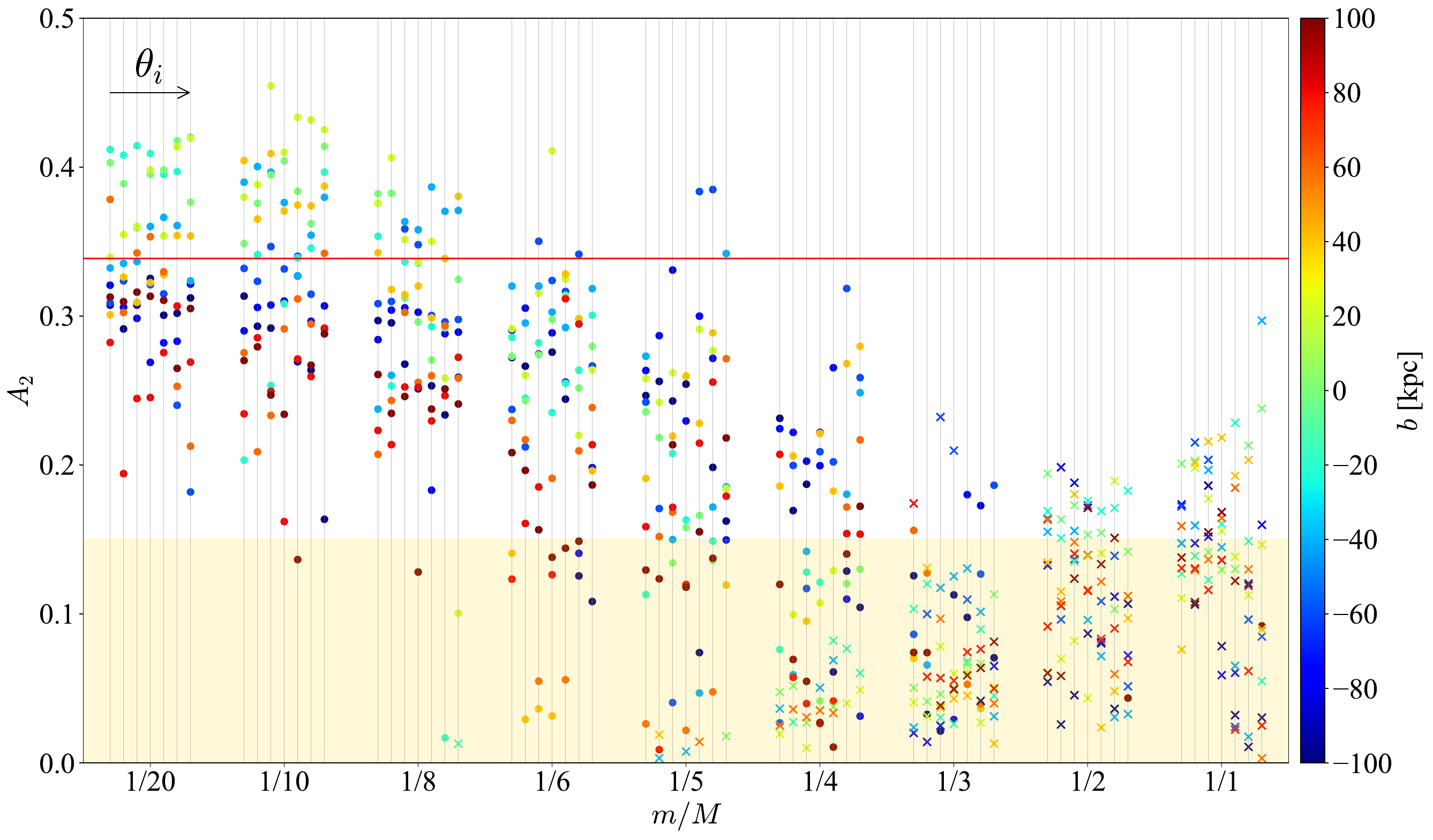}
    \caption{Distribution of final bar strength as a function of the
      merger parameters $(m/M,~b,~\theta_i)$. The mass ratio $m/M$ is
      shown on the horizontal axis. At each value of $m/M$, the seven
      vertical lines correspond to different inclination angles
      $\theta_i$, increasing from left to right ($\theta_i$ from
      $0^\circ$ to $90^\circ$). The color of each point indicates the
      impact parameter $b$. The yellow shaded region marks the weak-bar
      regime, defined as $A_2 < 0.15$. For comparison, the final bar
      strength of the isolated model is indicated by the red horizontal
      line. Models with negligible pattern speed
      ($\Omega_\mathrm{p}<0.25\,\Omega_\mathrm{p, isolated}$) are shown
    as crosses and are primarily located at higher $m/M$.}\label{fig: A2}
  \end{figure*}

  \begin{figure*}
    \centering
    \includegraphics[width=.98\textwidth]{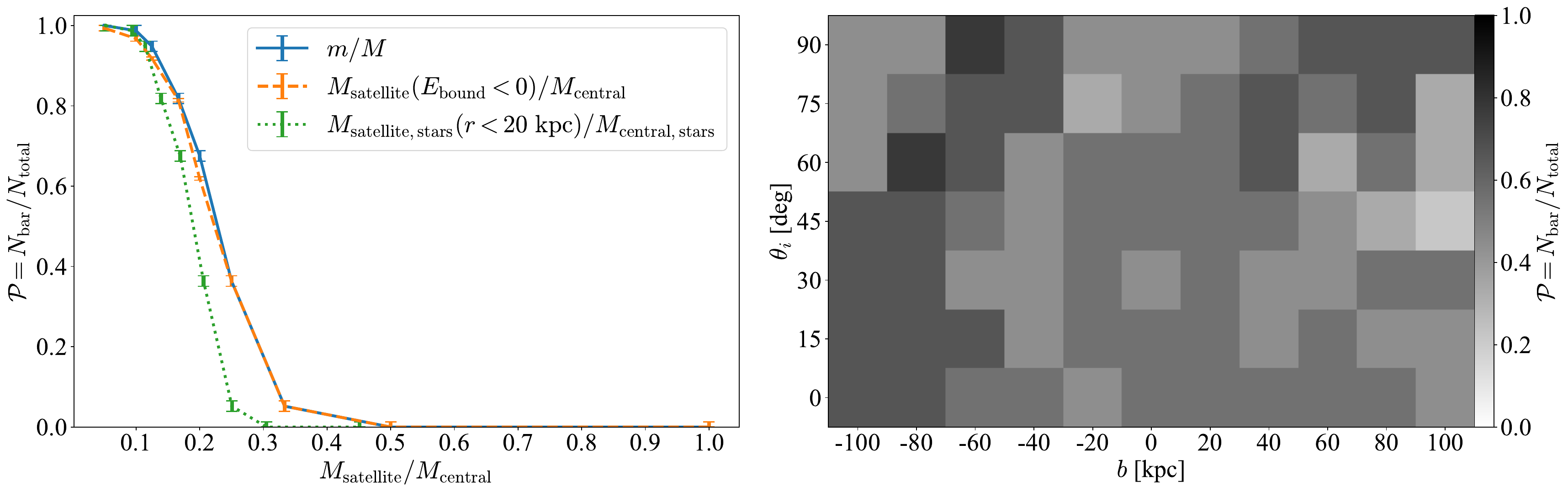}
    \caption{\textbf{Left}: Bar survival probability as a function of
      the satellite-to-central galaxy mass ratio, shown for three
      definitions: the initial mass ratio $m/M$ (blue), the bound mass
      of the satellite relative to the total mass of the central galaxy
      (orange), and the satellite's core stellar mass (within
      $20\ \mathrm{kpc}$ around its baryonic center) relative to the
      central galaxy's total stellar mass (green). In each mass ratio
      bin, $N_\mathrm{total}$ is the total number of models, and
      $N_\mathrm{bar}$ is the number of models hosting a bar at the end
      of the simulation. Vertical error bars indicate Poisson
      uncertainties in each bin. \textbf{Right}: Similar to the left
      panel, but showing the bar survival probability (indicated by the
      color of each pixel) as a function of $b$ and $\theta_i$. In each
      pixel, $\mathcal{P}$ is calculated among all $m/M$ models sharing
    the same $b$ and $\theta_i$.}\label{fig: bar survival}
  \end{figure*}

  \begin{figure*}
    \centering
    \includegraphics[width=.45\textwidth]{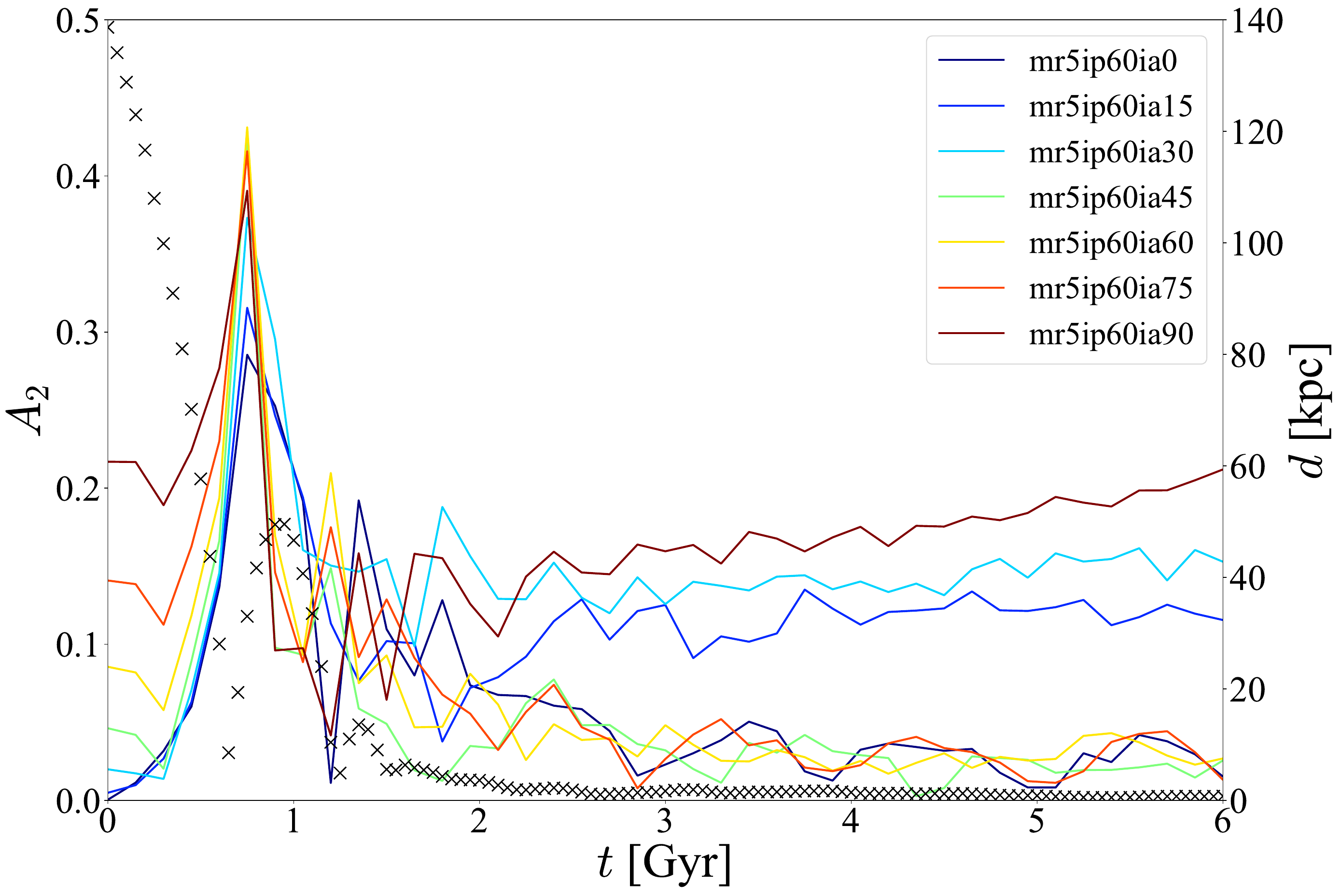}
    \hspace{1ex}
    \includegraphics[width=.38\textwidth]{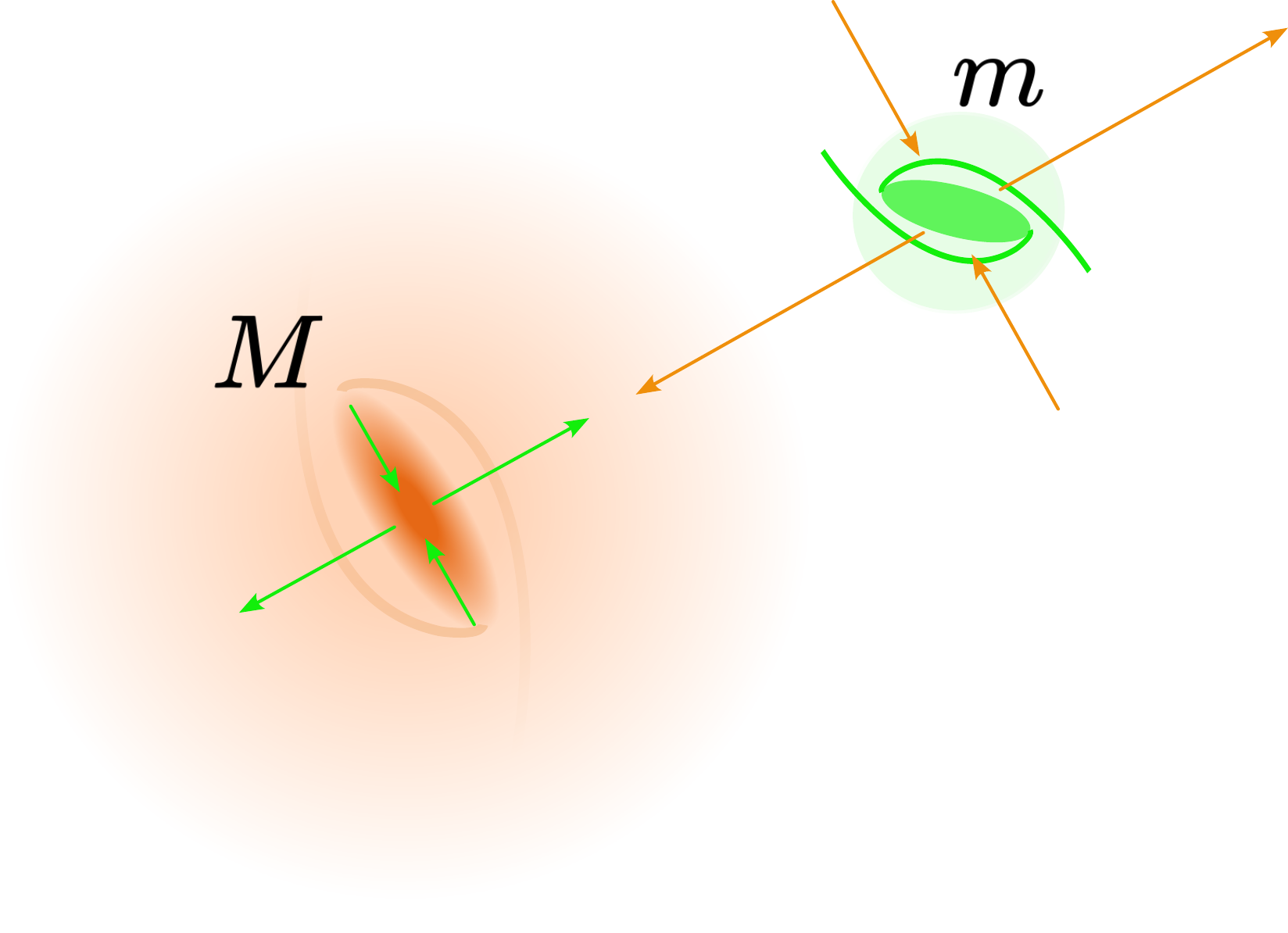}
    \includegraphics[height=.67\textheight]{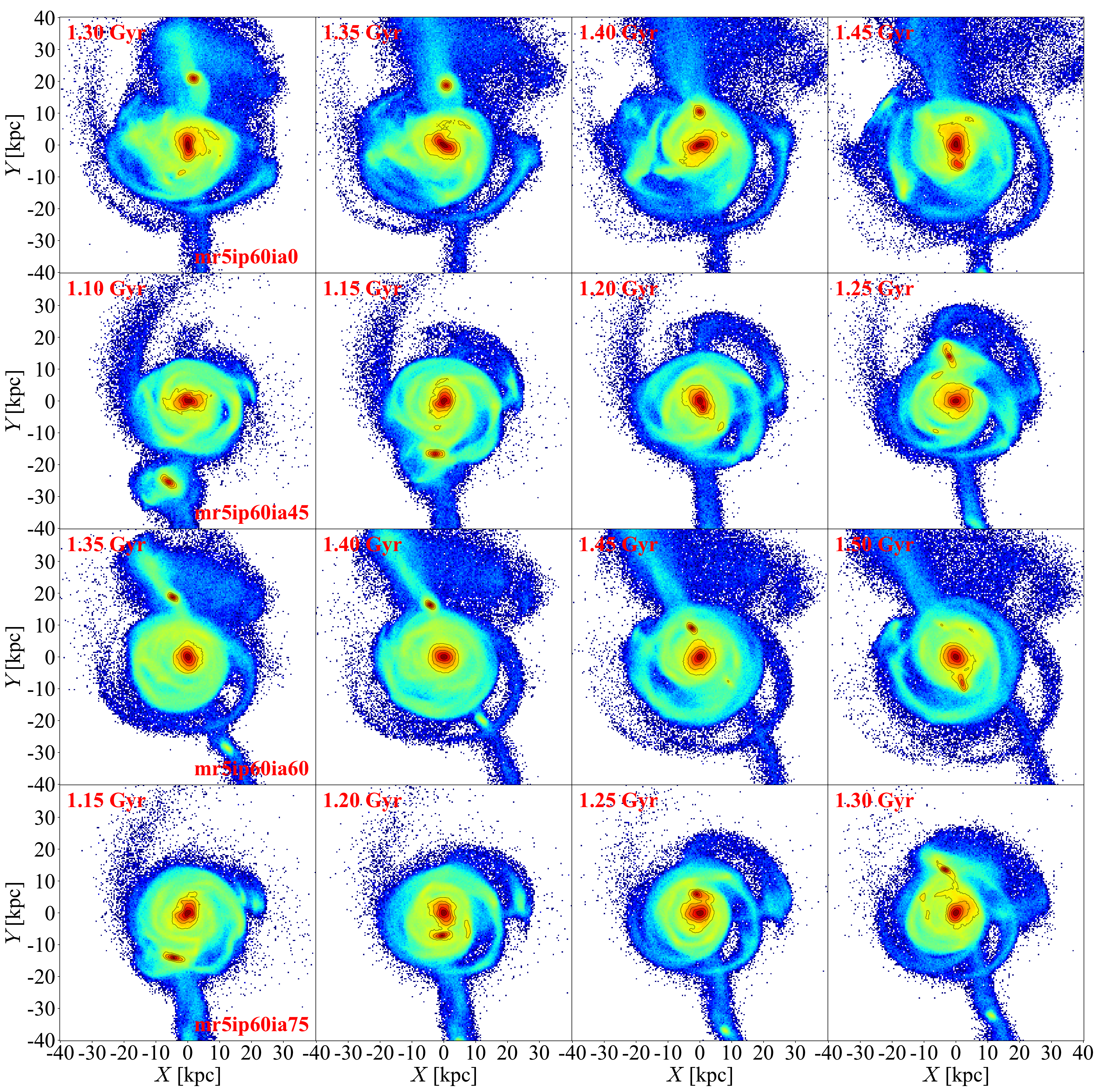}
    \caption{\textbf{Upper left}: Evolutionary histories of the bar
      strength $A_2(t)$ shown as solid lines and the distance $d(t)$
      between the galactic centers shown as crosses (averaged across
      these models). All models share the same merger mass ratio and
      impact parameter, but evolve toward distinct barred or unbarred
      morphologies by $6\ \mathrm{Gyr}$. \textbf{Lower rows}: Face-on
      views of the non-barred models in the \texttt{mr5ip60iaX}
      series. All exhibit severely deformed bars due to strong tidal
      disturbances from the satellites. \textbf{Upper right}:
      Schematic illustration of the tidal interaction between the
      central galaxy (orange) and the satellite (green).
    }\label{fig: demo series}
  \end{figure*}

  \section{Results}\label{sec: results}

  Figure~\ref{fig: snapshots} shows face-on views of the isolated
  model and several representative merger models at different epochs.
  In model \texttt{mr20ip-60ia45} (second row), the satellite is very
  light and thus exerts a negligible influence on the forming bar in
  the central galaxy. Consequently, the bar survives the merger, and
  its final morphology closely resembles that of the isolated model
  (top row). In model \texttt{mr5ip-60ia0} (third row), the bar is
  temporarily disturbed by the satellite but subsequently recovers.
  In contrast, in model \texttt{mr2ip100ia60} (fifth row), the
  satellite disrupts the forming bar, leaving behind an elliptical
  galaxy as expected for such a violent merger \citep{gerhar_1981,
  far_sha_1982, neg_whi_1983}. An intermediate case appears in model
  \texttt{mr5ip0ia90} (fourth row), where the bar initially survives
  the merger but gradually decays over time, ultimately evolving into
  a featureless disk. Notably, many of these merger models exhibit
  shell-like structures formed by ``phase wrapping'' during the
  merger process \citep[e.g.,][]{quinn_1984}.

  To quantify the presence of a bar, we measure the bar strength of
  each model at every snapshot. Because the satellite can tilt the
  disk of the central galaxy, we first realign the disk to its
  original plane before performing any bar-strength measurements. As
  illustrative examples, the bar strengths of the models in
  Figure~\ref{fig: snapshots} are annotated accordingly.

  We find that some models evolve into elliptical systems without a
  genuine bar, yet still show a prominent $A_2$ signal—such as model
  \texttt{mr2ip100ia60} in Figure~\ref{fig: snapshots}. A genuine bar
  exhibits significant pattern rotation, whereas in elliptical
  galaxies the apparent $m=2$ mode has nearly zero rotation. To
  differentiate between these cases, we also measure the bar pattern
  speed for each model. In elliptical systems, the bar pattern
  speed---or more strictly, the $m=2$ phase speed---is close to zero.

  Figure~\ref{fig: pattern speeds} shows the distribution of bar
  pattern speed as a function of bar strength at the end of the
  simulation ($10~\mathrm{Gyr}$). The models separate into two
  distinct groups: one with pattern speeds near that of the isolated
  model, $\Omega_\mathrm{p, isolated}$, and the other clustered
  around $\Omega_\mathrm{p}=0~\mathrm{km\,s^{-1}\,kpc^{-1}}$. Models
  with $\Omega_\mathrm{p}\simeq0$ are typically early-type systems
  characterized by a relatively low azimuthal streaming
  velocity–to–dispersion ratio (averaged over
  $10<R/\mathrm{kpc}<20$). Other models, particularly those with
  lower $m/M$, exhibit an anti-correlation between
  $\Omega_\mathrm{p}$ and $A_2$, reflecting the similar angular
  momentum in the bar region across models with similar merger orbits.

  To exclude non-rotating early-type systems, we define a model as
  hosting a \emph{genuine bar} at the end of the simulation if it
  satisfies: (1) $A_2 > 0.15$ and (2) $\Omega_\mathrm{p} >
  0.25\,\Omega_\mathrm{p, isolated}$, corresponding to the unshaded
  upper-right region of Figure~\ref{fig: pattern speeds}.

  Figure~\ref{fig: A2} presents the distribution of the final bar
  strength as a function of the merger mass ratio $m/M$, where models
  with $\Omega_\mathrm{p}<0.25\,\Omega_\mathrm{p, isolated}$ are
  marked as cross symbols. Overall, models with small satellites
  ($m/M \lesssim 1/10$) typically develop a prominent bar structure,
  with strengths comparable to or even exceeding that of the isolated
  model. As $m/M$ increases, the final $A_2$ of the models generally
  decreases. When $m/M$ reaches $1/4$, all models exhibit bar
  strengths lower than that of the isolated model. For even higher
  mass ratios, nearly all models display either a weak $A_2$ or
  negligible rotation (marked as crosses), which indicates that in
  these cases, the massive satellites---regardless of their $b$ or
  $\theta_i$--- violently disrupt the forming bar in the central galaxy.

  To better visualize the trend that bars are more likely to be
  destroyed by more massive satellites, the left panel of
  Figure~\ref{fig: bar survival} shows the bar survival probability,
  $\mathcal{P}$, as a function of
  $M_\mathrm{satellite}/M_\mathrm{central}$, where in each mass ratio bin
  \[ \mathcal{P} \equiv \frac{N_\mathrm{bar}}{N_\mathrm{total}}, \]
  with $N_\mathrm{total}$ the number of models and $N_\mathrm{bar}$
  the number of models hosting a bar at the end of the simulation.
  The blue points correspond to the initial mass ratio,
  $M_\mathrm{satellite}/M_\mathrm{central} = m/M$. Considering mass
  loss during the merger, the remaining satellite mass may be a more
  physically meaningful indicator than the initial mass. We estimate
  the ``remaining mass'' of the satellite using two definitions
  measured at the first apocenter of the merger: (1) the
  \textit{bound mass} (orange points) is the total mass of stellar
  and dark matter particles gravitationally bound to the satellite,
  i.e., particles with
  \[
    E_\mathrm{bound} = \Phi_\mathrm{satellite} +
    \frac{1}{2}v^2_\mathrm{in\,satellite} < 0,
  \]
  and (2) the \textit{core mass} (green points) is the stellar mass
  enclosed within $20\ \mathrm{kpc}$ of the satellite barycenter,
  where only stellar mass is used in calculating
  $M_\mathrm{satellite}/M_\mathrm{central}$. The left panel of
  Figure~\ref{fig: bar survival} demonstrates that the bar survival
  probability decreases systematically with increasing
  $M_\mathrm{satellite}/M_\mathrm{central}$ under all mass
  definitions. Nearly all bars survive mergers with
  $M_\mathrm{satellite}/M_\mathrm{central} \lesssim 0.1$, whereas
  almost all bars are destroyed when
  $M_\mathrm{satellite}/M_\mathrm{central} \gtrsim 0.4$, providing an
  upper limit on the possible merger mass ratio in the assembly
  history of a MW–like barred galaxy.

  Notably, compared with the merger mass ratio, the bar survival
  probability shows little dependence on $b$ or $\theta_i$ in
  Figure~\ref{fig: A2} (except for a weak secondary trend that, in
    models with $m/M \leq 1/8$, smaller $|b|$ is associated with
  slightly larger $A_2$). To better illustrate this, the right panel
  of Figure~\ref{fig: bar survival} presents the distribution of
  $\mathcal{P}$ as a function of $b$ and $\theta_i$, where
  $\mathcal{P}$ is calculated among models sharing the same $b$ and
  $\theta_i$. The nearly random distribution of $\mathcal{P}$ in the
  panel demonstrates that these parameters are not primary in shaping
  the bar survival probability compared with the merger mass ratio.
  We also examine alternative proxies for the impact parameter, such
  as orbital circularity (by \citealt{abadi_etal_2003}, 0–0.4 in our
  models) and ellipticity (0.7–1.0), and still find no significant
  dependence of bar survival on either proxy. Therefore, in our
  simulations, the bar survival probability is determined primarily
  by the merger mass ratio and shows no notable dependence on $b$,
  $\theta_i$, or related orbital parameters.

  \section{Discussion}\label{sec: discussion}

  \subsection{Influence of $b$ and $\theta_i$ on the forming bar}
  \label{sec: discussion: b and theta}

  Section~\ref{sec: results} shows that bar survival probability
  during mergers is governed primarily by the merger mass ratio. In
  the intermediate regime, $0.10\lesssim m/M \lesssim 0.40$, the
  survival probability lies between 0 and 1, indicating that bars
  survive in some mergers with particular combinations of $b$ and
  $\theta_i$ but are destroyed in others. This indicates that $b$ and
  $\theta_i$ play a secondary, albeit stochastic, role in determining
  whether a bar survives a given merger.

  For example, the upper left panel of Figure~\ref{fig: demo series}
  shows the bar evolutionary histories of models \texttt{mr5ip60iaX}
  along with $d(t)$ the distance between the galactic centers. These
  models share the same merger mass ratio and impact parameter, and
  thus follow similar merging orbits. Therefore, we plot only the
  averaged $d(t)$ across these models. They all exhibit two passages
  between $0\ \mathrm{Gyr}$ and $2\ \mathrm{Gyr}$ before
  transitioning into secular evolution as the intergalactic distance
  approaches zero. Despite their similar orbits, the models show
  remarkably different bar evolution: some develop prominent bars by
  the end, while others do not. Given the similar merger orbits, this
  divergence in bar evolution must arise from some subtle differences
  in the merger process that ultimately shape the final barred or
  unbarred morphologies.

  To identify the subtle factors responsible for the divergent bar
  evolution, the lower rows of Figure~\ref{fig: demo series} present
  several key snapshots of the non-barred models identified in the
  upper left panel, tracing the bar destruction process in each case.
  The snapshots show that the satellites cross the forming bars along
  directions nearly perpendicular to the bar's major axis at certain
  times (possibly between the displayed snapshots). During these
  passages, the satellites induce strong disturbances, evident in the
  significantly distorted bars, which ultimately deconstruct the
  forming bars in these models.

  The tidal forces exerted by the satellite on the primary galaxy and
  the forming bar can account for such violent disturbances. The
  upper right panel of Figure~\ref{fig: demo series} illustrates the
  physical impact of the tidal force during a merger. The satellite
  exerts an extensive tidal force along the axis connecting the
  galactic centers, while a compressive force acts in directions
  perpendicular to it. If the bar's major axis is aligned
  perpendicular to this axis, it experiences compression along its
  major axis and extension along the orthogonal directions, tending
  to circularize the structure. This expectation is consistent with
  the morphological distortions shown in the lower rows. When the
  infalling satellite passes near the bar's minor axis, it can
  distort the forming bar into a rounder shape. Thus, a satellite on
  an orbit nearly perpendicular to the bar can exert a tidal force
  capable of deconstructing it. Whether the bar is ultimately
  destroyed depends on the tidal force strength,
  $F_\mathrm{tidal}\propto m/d^3$, time span of the nearly
  perpendicular passage, and the robustness of the bar structure
  against external perturbations. Figure~\ref{fig: demo series} shows
  cases in which the satellite's tidal influence is strong enough to
  destroy the forming bar.

  This physical picture explains the dependency of bar survival
  probability on the three merger parameters. The influence of $b$
  and $\theta_i$ is stochastic: the tidal force disrupts the bar only
  when the satellite happens to align nearly perpendicular to the
  bar’s major axis for a sufficient duration while crossing the bar.
  Even in such cases, only sufficiently massive satellites can exert
  a strong enough tidal perturbation to disrupt the bar as the tidal
  force scales as $F_\mathrm{tidal} \propto m$. For a tiny satellite,
  with $m/M\lesssim1/10$, almost every bar can survive the merger. As
  the satellite mass increases, both the strength and the spatial and
  temporal extent of the bar-deconstructive tidal force grow, making
  bar destruction more likely during the merger. With even higher
  satellite mass, for $m/M \geq 1/2$, even the disk can be strongly
  distorted or destroyed, explaining why nearly all bars are
  disrupted in such models.

  Notably, the above physical framework is purely dynamical and does
  not depend on whether the merger occurs during or after bar
  formation, nor on the presence of a gas component. The only
  requirement is that the central galaxy hosts a (formed or forming)
  bar pattern capable of interacting with the merging satellite.
  Consequently, this explanation should also apply to
  post-bar-formation merger events (with or without gas), where we
  expect a similar dependence of the bar survival probability on
  $(m/M,~b,~\theta_i)$. For example, with post-bar-formation mergers
  that include a non-trivial gas component, \citet{zhou_etal_2025}
  demonstrate qualitatively similar results.

  \begin{figure*}
    \centering
    \includegraphics[width=.98\textwidth]{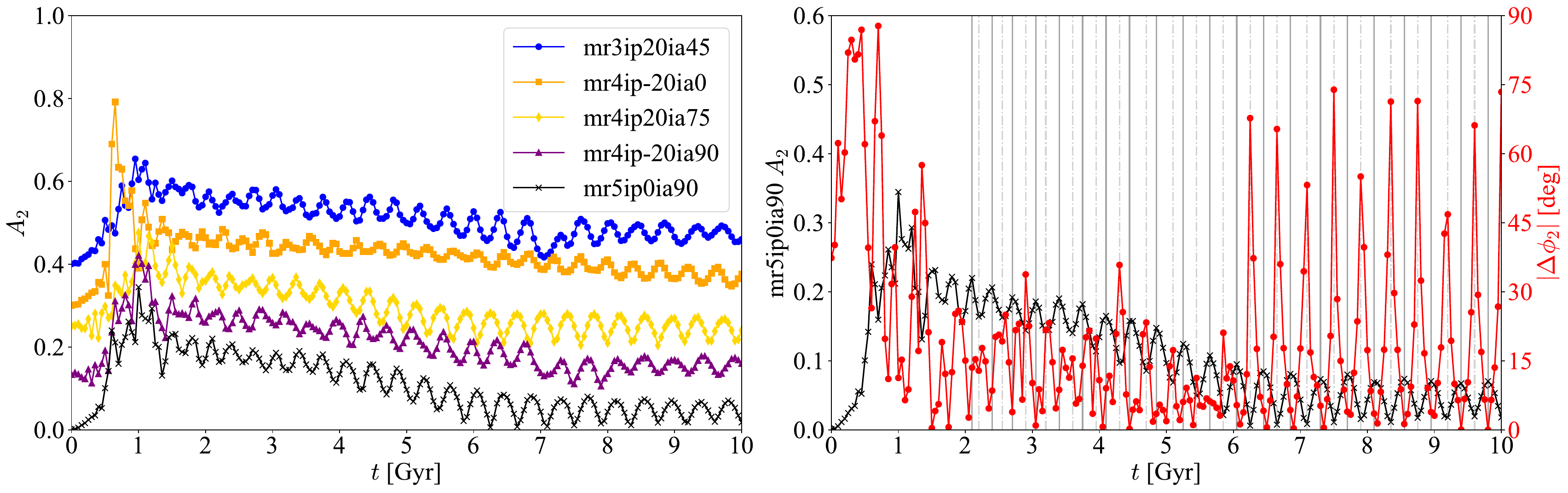}
    \caption{\textbf{Left}: $A_2(t)$ curves of the weakening bars.
      For clarity, the curves are vertically offset by $0.1$
      \textbf{from each other}. \textbf{Right}: Evolution of $A_2$ in
      model \texttt{mr5ip0ia90}, alongside $m=2$ phase deviations
      between the inner ($R<7.5\ \mathrm{kpc}$) and outer
      ($7.5\ \mathrm{kpc}<R<15\ \mathrm{kpc}$) bar regions. After the
      merger ($t\gtrsim2\ \mathrm{Gyr}$), the maxima and minima of
      $A_2$ are marked by vertical gray solid and dashed lines,
      respectively. The minima of $A_2$ align with maxima of
      $|\Delta\phi_2|$ approximately, indicating a tight correlation
      between bar strength and double-bar interaction.
    }\label{fig: weakening bars}
  \end{figure*}

  \subsection{Weakening bars}\label{sec: discussion: weakening}

  Figure~\ref{fig: snapshots} shows that the merger event in model
  \texttt{mr5ip0ia90} does not completely destroy the bar. However,
  the bar gradually weakens after the merger and eventually
  disappears. A similar phenomenon, referred to as a bar-weakening
  event, has been reported in minor mergers by
  \citet{ghosh_etal_2021}. Here, we define a (significantly)
  weakening bar based on the following criteria:
  \begin{enumerate}[itemsep=0pt, topsep=0pt, parsep=0pt]
    \item[(a)] Right after the merger, the system hosts a prominent
      bar: $A_\mathrm{2,m} > 0.15$, where $A_\mathrm{2,m} \equiv
      \operatorname{mean}\{A_2(t):2\ \mathrm{Gyr}<t<3\ \mathrm{Gyr}\}$.
    \item[(b)] The final bar strength is significantly decreased
      relative to the value right after the merger:
      $A_\mathrm{2,f}<0.5\, A_\mathrm{2,m}$.
    \item[(c)] The final snapshot lacks a prominent bar: $A_\mathrm{2,f}<0.15$.
  \end{enumerate}

  Among the 693 merger models, we identify 5 cases exhibiting
  significant bar weakening, with their $A_2(t)$ curves shown in the
  left panel of Figure~\ref{fig: weakening bars}. An additional 8
  models show milder weakening if we relax condition (b) from $0.5\,
  A_\mathrm{2,m}$ to $0.75\, A_\mathrm{2,m}$ and remove condition
  (c). They account for approximately $0.7\%$ and $1\%$ of all
  models, respectively, indicating that bar weakening is a rare
  phenomenon. All identified bar weakening models fall within the
  range $1/5 \leq m/M \leq 1/3$, corresponding to the intermediate
  regime in the probability curves of Figure~\ref{fig: bar survival}.

  A possible cause of the weakening bars is the interaction between
  the nested double bars in these models. As shown in the left panel
  of Figure~\ref{fig: weakening bars}, in addition to the secular
  decline in bar strength after the merger, the bar strength exhibits
  periodic oscillations. Inspection of the face-on views reveals that
  these models develop double-bar features, formed from clumpy merger
  debris (rather than internal dynamical instabilities as in
    \citealt{deb_she_2007}, \citealt{she_deb_2009}, and
  \citealt{du_etal_2015}). For example, model \texttt{mr5ip0ia90},
  quantified by the right panel of Figure~\ref{fig: weakening bars},
  presents the phase deviation $|\Delta\phi_2|$ between the inner ($R
  < 7.5~\mathrm{kpc}$) and outer ($7.5~\mathrm{kpc} < R <
  15~\mathrm{kpc}$) bar regions showing a significant offset, which
  indicates the presence of a double-bar pattern. Notably, although
  with some noise, the bar strength $A_2(t)$ usually reaches minima
  when the phase deviation reaches maxima. Similar behaviors are seen
  in other bar weakening models, suggesting a strong correlation
  between bar strength evolution and the relative phase of the nested
  bars. This phenomenon resembles the double-bar interaction
  described by \citet{deb_she_2007} and \citet{she_deb_2009}, who
  directly constructed isolated double-barred galaxy models. In their
  run 1, a similar bar weakening process was observed: the dominant
  bar pattern gradually decreased from above $0.25$ to below $0.1$,
  also accompanied by periodic oscillations. Oscillations in bar
  strength were also notable in the double-bar models of
  \citet{du_etal_2015} and the weakening bar models of
  \citet{ghosh_etal_2021}. These similarities suggest that the
  weakening is likely driven by dynamical interaction between the
  inner and outer bars. Thus, phenomenologically, the coupled
  evolution of nested bars appears to be responsible for both the
  oscillations and the secular decline in bar strength in these models.

  Other processes, such as central mass concentration and angular
  momentum exchange discussed by \citet{ghosh_etal_2021}, may also
  contribute to bar weakening. However, given the similarity in
  frequency between the rapid oscillations of the $A_2(t)$ curves and
  the double-bar phase deviations, the weakening bars in our models
  appear more consistent with mutual interactions between nested double bars.

  The interaction between nested bars provides a natural explanation
  for the low probability of forming bar weakening models. Such
  interactions arise only when a merger event coincidentally produces
  a double-bar system, allowing the inner and outer bars to interact
  and undermine the global bar strength. This scenario also explains
  why bar weakening models occur within the merger mass-ratio range
  $1/5 \leq m/M \leq 1/3$: if the satellite is too massive, the
  forming bar is directly disrupted by the merger; if it is too
  small, it cannot produce a prominent double-bar pattern.
  Consequently, bar weakening is confined to the transitional regime
  of the bar survival probability curves shown in
  Figure~\ref{fig: bar survival}.

  Note that while interactions between nested bars provide a
  plausible mechanism for bar weakening, they do not always lead to a
  decline in bar strength. For example, in run 2 of
  \citet{deb_she_2007} and the clumpy double-bar model of
  \citet{du_etal_2015}, the nested bars induce periodic oscillations
  in bar strength without significantly weakening the global bar
  strength. The conditions under which such interactions can weaken a
  bar remain to be explored. A detailed investigation of this topic
  is beyond the scope of the present work and will be addressed in a
  follow-up paper.

  In summary, the bar weakening events identified in our models are
  likely driven by interactions between nested bars, which occur
  occasionally in the transitional regime of the bar survival probability curve.

  \section{Summary}\label{sec: summary}

  We have constructed 693 \textit{N}-body models of GSE-like radially
  baised major merger events, each consisting of a central and a
  satellite disk galaxy, with varying merger mass ratios $m/M$,
  orbital impact parameters $b$, and orbital inclination angles
  $\theta_i$. The satellite's initial conditions are set up so that
  it merges into the central galaxy around $1.5\ \mathrm{Gyr}$,
  during the spontaneous bar formation process of the central galaxy.

  We find that the bar survival probability decreases primarily with
  increasing $m/M$ (Figure~\ref{fig: bar survival}). Bars nearly
  always survive mergers with $m/M \lesssim 0.1$, but are never
  preserved in mergers with $m/M \geq 0.5$, where the galaxies evolve
  into more early-type-like systems that may appear as non-rotating
  bar-like ellipticals (Figures~\ref{fig: snapshots}
  and \ref{fig: pattern speeds}). Within the transitional regime
  between definite bar survival and destruction
  ($1/5 \leq m/M \leq 1/3$), several models show bar weakening phenomenon
  (Figure~\ref{fig: weakening bars}), possibly
  driven by interactions between nested double bars
  formed from merger debris (Section~\ref{sec: discussion: weakening}).

  In contrast, the impact parameter $b$ and inclination angle
  $\theta_i$ exert secondary and stochastic effects on bar survival
  compared with $m/M$. The tidal force exerted by the satellite on
  the forming bar naturally explains the dependencies of bar survival
  probability on these three merger parameters
  (Section~\ref{sec: discussion: b and theta}).
  This tidal interaction is purely
  dynamical and therefore independent of whether it occurs
  during-bar-formation or post-bar-formation, or of the presence of
  gas content, implying that post-bar-formation and gas-rich mergers
  may also follow similar behavior.

  \begin{acknowledgments}

    We thank the anonymous referee for the careful review. We
    sincerely thank Victor P. Debattista, Zhao-Yu Li, and Hui Li for
    helpful discussions. Bin-Hui Chen gratefully acknowledges the
    financial support from the China Scholarship Council, the support
    of the T.D. Lee Scholarship, and the sponsorship from Yangyang
    Development Fund. The research presented here is partially
    supported by the National Natural Science Foundation of China
    under grant Nos. 12533004, 12025302, 11773052; by China Manned
    Space Program with grant no. CMS-CSST-2025-A11; by the ``111''
    Project of the Ministry of Education of China under grant No.
    B20019; and by Office of Science and Technology, Shanghai
    Municipal Government with grant Nos. 24DX1400100 and
    ZJ2023-ZD-001. This work used the Gravity Supercomputer at the
    Department of Astronomy, Shanghai Jiao Tong University, and the
    facilities in the National Supercomputing Center in Jinan.

    \textit{Software}: \texttt{AGAMA} \citep{vasili_2019},
    \texttt{GADGET-4} \citep{spring_2005, spring_etal_2021},
    \texttt{Numpy} \citep{numpy}, \texttt{Scipy} \citep{scipy},
    \texttt{Matplotlib} \citep{matplotlib}, and
    \href{https://inkscape.org/}{\texttt{Inkscape}}.

  \end{acknowledgments}

  \bibliography{references}{}
  \bibliographystyle{aasjournalv7}

  \end{document}